\documentclass[lettersize,journal]{IEEEtran}
\usepackage{amsmath,amsfonts}
\usepackage{algorithmic}
\usepackage{algorithm}
\usepackage{array}
\usepackage[caption=false,font=footnotesize,labelfont=rm,textfont=rm]{subfig}
\usepackage{titlesec}
\usepackage{color}
\usepackage{graphicx}
\usepackage{textcomp}
\usepackage{stfloats}
\usepackage{url}
\usepackage{verbatim}
\usepackage{graphicx}
\usepackage{epstopdf}
\usepackage{cite}
\begin{document}

\title{
	\huge{Message Feedback Interference Cancellation Aided UAMP Iterative Detector for OTFS Systems}
}

\author{Xiangxiang Li, Haiyan Wang, \emph{Member, IEEE}, Yao Ge, \emph{Member, IEEE}, \\ Xiaohong Shen, \emph{Member, IEEE} and Jiarui Zhao
	
	\vspace{-10pt}
	\thanks{ This work was supported by the NSFC project No. 62031021. 
	The work of Yao Ge was supported under the RIE2020 Industry Alignment Fund-Industry Collaboration Projects (IAF-ICP) Funding Initiative, as well as cash and in-kind contribution from the industry partner(s).
	(\emph{Corresponding author: Haiyan Wang}).
	}
	
	\thanks{Xiangxiang Li, Haiyan Wang, Xiaohong Shen and Jiarui Zhao are with the School of Marine Science and Technology and the	Key Laboratory of Ocean Acoustics and Sensing, Ministry of Industry and Information Technology, Northwestern Polytechnical University, Xi’an, Shaanxi 710072, China. Haiyan Wang is also with the School of Electronic Information and Artifcial Intelligence, Shaanxi University of Science and Technology,	Xi’an, Shaanxi 710021, China (e-mail: lixx@mail.nwpu.edu.cn; hywang@nwpu.edu.cn; xhshen@nwpu.edu.cn; zjr@mail.nwpu.edu.cn).}
	\thanks{Yao Ge is with the Continental-NTU Corporate Laboratory, Nanyang Technological University, Singapore 637553 (e-mail: yao.ge@ntu.edu.sg).}
	\vspace{-20pt}
}

\maketitle
\setlength{\textfloatsep}{4.5pt}

\begin{abstract}

The designing of efficient signal detectors is important and yet challenge for orthogonal time frequency space (OTFS) systems in high-mobility scenarios. In this letter, we develop an efficient message feedback interference cancellation aided unitary approximate message passing (denoted as UAMP-MFIC) iterative detector, where the latest feedback messages from variable nodes are utilized for more reliable interference cancellation and performance improvement. A fast recursive scheme is leveraged in the proposed UAMP-MFIC detector to prevent complexity increasing. To further alleviate the error-propagation and improve the receiver performance, we also develop the bidirectional symbol detection structures, where Turbo UAMP-MFIC detector and iterative weight UAMP-MFIC detector are proposed to efficiently fuse the estimation results of forward and backward UAMP-MFIC detectors. The simulation results are finally provided to demonstrate performance improvement of our proposed detectors over existing detectors. 

\end{abstract}


\vspace{-11pt}
\section{Introduction}
\IEEEPARstart {T}{he} orthogonal time frequency space (OTFS) modulation has shown stronger robustness than widely used orthogonal frequency division multiplexing (OFDM) modulation in high-mobility scenarios. Different from OFDM, OTFS maps the  information symbols into delay-Doppler (DD) domain rather than traditional time-frequency (TF) domain, which can effectively utilize both delay and Doppler channel diversity for performance improvement. In order to harvest the potential channel diversity of OTFS system, the efficient and robust detection algorithms for equalization with interference cancellation are important and challenging.

The joint maximum a \emph{posteriori} probability (MAP) detector can achieve optimal detection performance, but it is intractable in practice due to the complexity growing exponentially with the system dimension\cite{c14}. The classic linear detectors such as zero-forcing (ZF) and linear minimum mean square error (LMMSE) \cite{c1} suffer from severe performance-loss and high complexity due to large dimension matrix inverse. Although the OFDM-based OTFS structure \cite{c2} can block diagonalize time-domain channel matrix by inserting a cyclic prefix (CP) in every OFDM symbol of each OTFS frame to reduce the dimension of matrix inverse, it results in a low spectral efficiency. Therefore, the limitations of linear detectors motivate the employment of non-linear detectors for OTFS systems.

The soft decision feedback equalizer (SDFE) \cite{c3} can effectively equalize the DD channels by non-linear structure with feed-forward and feedback filtering, but it still bears the high computational load from large dimension matrix inverse in LMMSE feed-forward filtering.
The Gaussian message passing (GMP) \cite{c4}\cite{c5} and expectation propagation (EP) \cite{c6}\cite{c7} detectors are proposed for signal detection with a low complexity based on sparse DD domain channel matrix, but the detection performance may degrades significantly when so many short loops exist in the dense factor graph. The approximate message passing (AMP) detector \cite{c8} is proposed based on loopy belief propagation with proper approximations, but its performance is limited to the large i.i.d. Gaussian channel matrix. To tackle this issue, the unitary AMP (UAMP) detector \cite{c13}\cite{c10} can effectively address the limitations of AMP detector for specific channel matrix and ensure the convergence performance by performing the unitary transform for original channel matrix. However, it is worth noting that the original UAMP detector adopts the parallel processing scheme at variable nodes and can not exploit latest feedback messages for interference cancellation, leading to performance loss and unsatisfied convergence rate.

To further accelerate the convergence and achieve performance improvement, we propose a message feedback interference cancellation aided UAMP (denoted as UAMP-MFIC) iterative detector in this letter, where the latest feedback messages from variable nodes are utilized for more reliable interference cancellation. To alleviate the error-propagation and further improve the detection performance, we develop the bidirectional symbol detection structures, where Turbo UAMP-MFIC (T-UAMP-MFIC) detector and iterative weight UAMP-MFIC (IW-UAMP-MFIC) detector are proposed to efficiently fuse the estimation results of forward and backward UAMP-MFIC detectors. In particular, the proposed IW-UAMP-MFIC detector shows better convergence performance but slower convergence rate than T-UAMP-MFIC detector. The simulation results are finally provided to demonstrate performance improvement of the proposed detectors over existing detectors. It is worth mentioning that our proposed message feedback interference cancellation structure can be also extended to the general MP-type detectors, such as GMP \cite{c4}\cite{c5}, EP \cite{c6}\cite{c7}, AMP \cite{c8} and so on.

\section{System Model}

In this section, we briefly outline the system model and the  input-output relationship of OTFS system in DD domain. Without loss of generality, we consider the $ N $ time slots and $ M $ sub-carriers for each OTFS frame. Specifically,  the information sequences $\mathbf{x} \in \mathbb{A}^{MN \times 1} $ are arranged into the two-dimensional matrix $\mathbf{X}_{\text{DD}} \in \mathbb{A}^{M \times N} $ and map into the DD domain grids, i.e. $\mathbf{X}_\textnormal{DD} = \mathbf{invec} \left( \mathbf{x} \right) $, where $\mathbb{A} $ is the normalized constellation set. Then the time domain signals $\mathbf{x}_{\text{T}} \in \mathbb{C}^{MN \times 1}$ can be obtained by firstly applying the inverse symplectic finite Fourier transform (ISFFT) and then Heisenberg transform to $\mathbf{X}_\textnormal{DD}$ with a rectangular pulse,
\begin{equation}\label{1}
	\mathbf{x}_{\text{T}} = \mathbf{vec} \left\lbrace  \mathbf{F}^\textnormal{H}_M \left( \mathbf{F}_M \mathbf{X}_\textnormal{DD} \mathbf{F}^\textnormal{H}_N \right)  \right\rbrace = \left( \mathbf{F}^\textnormal{H}_N \otimes \mathbf{I}_{\text{M}} \right) \mathbf{x},
\end{equation}
where $\mathbb{C}$ and $\otimes$ denote the complex number field and Kronecker product operator, respectively. $ \mathbf{F}_M \in \mathbb{C}^{M \times M} $ and $ \mathbf{F}_N \in \mathbb{C}^{N \times N}$ represent the normalized $ M $-point and $ N $-point discrete fast Fourier transform (FFT) matrices, respectively. 

We append a CP of length no shorter than the maximal channel delay spread for whole OTFS frame to overcome the inter-frame interference. The time domain effective channel matrix $\mathbf{H}_{\textnormal{T}} \in \mathbb{C}^{MN \times MN} $ can be given by
\begin{equation}\label{2}
	\mathbf{H}_{\textnormal{T}} = \sum_{i=1}^{P} h_i \boldsymbol{\Pi}^{l_i} \boldsymbol{\Delta}^{k_i + \kappa_i},
\end{equation}
where $P$ is the number of propagation paths. The $h_i$, $l_i$, $k_i$ and $\kappa_i$ represent the $i$-th path complex gain, delay, integer and fractional Doppler shift, respectively. The $\boldsymbol{\Pi}$ is a forward cyclic shift matrix, 
\begin{equation}\label{3.1}
	\boldsymbol{\Pi} =
	\left[ 
	\begin{array}{cccc}
		0 & \cdots & 0 & 1 \\
		1 & \ddots & 0 & 0 \\
		\vdots & \ddots & \ddots & \vdots \\
		0 & \cdots & 1 & 0
	\end{array}
	\right]_{MN \times MN} ,
\end{equation}
and $\boldsymbol{\Delta} = \text{diag} \left[ 1, e^{j2\pi \frac{1}{MN}}, \ldots, e^{j2\pi \frac{MN-1}{MN} } \right]$ is a diagonal matrix. The received time domain signals $ \mathbf{y}_{\text{T}} \in \mathbb{C}^{MN \times 1} $ after removing CP can be expressed as
\begin{equation}\label{3}
	\mathbf{y}_{\text{T}} = \mathbf{H}_{\text{T}} \mathbf{x}_{\text{T}} + \boldsymbol{\omega}_{\text{T}},
\end{equation}
where $\boldsymbol{\omega}_{\text{T}} \sim  \mathcal {CN} (\boldsymbol{\omega}_{\text{T}};\boldsymbol{0},\gamma\boldsymbol{1})$ is additive white Gaussian noise (AWGN) with zero-mean and variance $\gamma$.

The corresponding DD domain received signals $\mathbf{y} \in \mathbb{C}^{MN \times 1}$ can be obtained by firstly performing the Wigner transform with a rectangular pulse and then symplectic finite Fourier transform (SFFT), i.e.
\begin{equation}\label{4}
	\mathbf{y} = \mathbf{vec} \left\lbrace \mathbf{F}^\textnormal{H}_M \left(  \mathbf{F}_{\text{M}} \mathbf{Y}_\textnormal{T} \right)  \mathbf{F}_N \right\rbrace = \left( \mathbf{F}_{\text{N}} \otimes \mathbf{I}_{\text{M}} \right) \mathbf{y}_{\text{T}},
\end{equation}
where $\mathbf{Y}_\textnormal{T} = \textbf{invec} \left( \mathbf{y}_{\text{T}} \right) $. Therefore, the input-output relationship of OTFS system in DD domain can be expressed as
\begin{equation}\label{5}
	\mathbf{y} =  \mathbf{H}_{\text{DD}} \mathbf{x} + {\boldsymbol{\omega}}_{\text{DD}},
\end{equation}
where $	\mathbf{H}_{\text{DD}} = \left( \mathbf{F}_{\text{N}} \otimes \mathbf{I}_{\text{M}} \right)  \mathbf{H}_{\text{T}} \left( \mathbf{F}_{\text{N}}^{\text{H}} \otimes \mathbf{I}_{\text{M}} \right)$ is the effective DD domain channel matrix and ${\boldsymbol{\omega}}_{\text{DD}} = \left( \mathbf{F}_{\text{N}} \otimes \mathbf{I}_{\text{M}} \right) \boldsymbol{\omega}_{\text{T}}$ is the effective DD domain noise with same distribution as $ \boldsymbol{\omega}_{\text{T}} $.

According to \eqref{5}, we can observe that it is challenging to recover transmitted signals $\mathbf{x}$ from received signals $\mathbf{y}$ directly due to the complex channel matrix structure $\mathbf{H}_{\text{DD}}$. Therefore, developing efficient and robust signal detectors are rather important and yet challenge. 

\section{Proposed UAMP-MFIC Iterative detector}

\begin{figure}[htbp]
	\centering
	\includegraphics[scale=0.45]{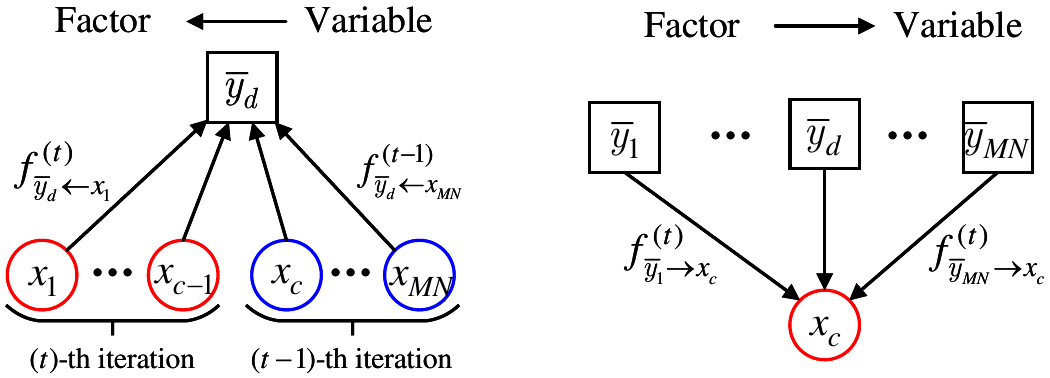}
	\caption{UAMP-MFIC detection for $c$-th symbol in $\left( t \right) $-th iteration.}
	\label{fig_1}
\end{figure}

The UAMP iterative detector firstly performs singular value decomposition (SVD) for $\mathbf{H}_{\text{DD}}$, i.e. $\mathbf{H}_{\text{DD}} = \mathbf{U} \boldsymbol{\Lambda} \mathbf{V}$, where $\mathbf{U}$ and $\mathbf{V}$ are the unitary matrices, $\boldsymbol{\Lambda}$ is the diagonal matrix consisted of the singular values of matrix $\mathbf{H}_{\text{DD}} $. Then the input-output relationship in \eqref{5} can be rewritten as
\begin{equation}\label{7}
	\bar{\mathbf{y}} =  \mathbf{H} \mathbf{x} + {\boldsymbol{\omega}},
\end{equation}
where $\bar{\mathbf{y}} =  \mathbf{U}^{\text{H}} \mathbf{y}$, $\mathbf{H} = \mathbf{\Lambda} \mathbf{V}$, $\boldsymbol{\omega} = \mathbf{U}^{\text{H}} \boldsymbol{\omega}_{\text{DD}}$ is the noise with same distribution as $\boldsymbol{\omega}_{\text{DD}}$ since $\mathbf{U}$ is the unitary matrix. Therefore, we aim to recover the original transmitted signals $\mathbf{x}$ from $\bar{\mathbf{y}}$ given matrix $\mathbf{H}$. The optimal MAP detector is intractable in practice due to high computational load. Then, we apply a sub-optimal symbol-by-symbol MAP detection scheme and the $c$-th estimation symbol $x_c$ can be expressed as
\begin{align}
		\hat{x}_c & = \mathop{\arg \max}\limits_{\alpha \in \mathbb{A}}
		\Pr \left( x_c = \alpha \left| \bar{\mathbf{y}}, \mathbf{H}\right. \right) \nonumber \\
		& \propto \mathop{\arg \max}\limits_{\alpha \in \mathbb{A}}  \Pr \left( \bar{\mathbf{y}} \left| x_c = \alpha,  \mathbf{H}\right. \right) P_{D} \left(x_c = \alpha   \right) \nonumber \\
		&\propto  \mathop{\arg \max}\limits_{\alpha \in \mathbb{A}} \prod_{d = 1 }^{MN} \Pr \left(\bar{y}_d \left. \right| x_c = \alpha , \mathbf{h}_{d} \right) P_{D} \left(x_c = \alpha   \right), \label{9.1}
\end{align}
where $\mathbf{h}_d$ is the $d$-th row of matrix $\mathbf{H}$, the priori probability $ P_{D} $ is assumed to be a Gaussian distribution, modeled as $\mathcal{CN} \left( x_c; \hat{\mu}_c^{\left( 0 \right)}, \hat{\eta}_c^{\left( 0 \right) } \right) $. Then we can use a factor graph to describe the model \eqref{9.1} in Fig. \ref{fig_1}, where factor nodes $\bar{y}_d, d = 1, 2, ..., MN $ and variable nodes $x_c, c = 1, 2, ..., MN$ are connected to each other based on matrix $\mathbf{H}$. 

To further enhance the interference cancellation and achieve performance improvement, we develop a message feedback interference cancellation structure and propose the UAMP-MFIC iterative detector, where the latest feedback messages from variable nodes are utilized for more reliable interference cancellation. Specifically, for $c$-th symbol in $\left( t \right) $-th iteration, the latest feedback extrinsic message $ f^{\left( t \right) }_{\bar{y}_d \leftarrow x_l}, l = 1, 2, ..., c-1 $ from variable node $x_l$ to factor node $\bar{y}_d $ substitutes the priori message $ f^{\left( t-1 \right) }_{\bar{y}_d \leftarrow x_l} $ in original UAMP detector for more reliable interference cancellation and performance improvement. In addition, the fast recursive scheme is leveraged in the proposed UAMP-MFIC detector to prevent complexity increasing. By exchanging the message between factor nodes and variable nodes iteratively, we can approximate \emph{a posteriori} distribution of detection symbols with faster convergence and better performance. The proposed UAMP-MFIC iterative detector can be summarized in \textbf{Algorithm \ref{alg1}} and we can describe the specific details in $\left( t \right) $-th iteration as following. 

\begin{algorithm}[t!]
	\caption{UAMP-MFIC Iterative Detector }\label{alg1}
	Input: ${\bar{\mathbf{y}}}$, $\mathbf{H}$,  $n_{iter}$, $ \mathbf{P}_{D} \left( x_c \right), c = 1, 2, ..., MN.  $\\
	Initialize: $\boldsymbol{\mu}^{\left( 0 \right) } $, $\boldsymbol{\eta}^{\left( 0 \right) }$ and $t = 1$. \\
	\textbf{Repeat}
	\begin{algorithmic}[1]
		\FOR {$ c=1 $ to $ MN $}
		\STATE Factor node $\bar{y}_d$ generates mean $\chi_{d,c}^{\left( t \right) }$ and variance $\tau_{d,c}^{\left( t \right) }$ in \eqref{13}, then passes them to the variable node $x_c$, $d = 1, 2, ..., MN$;
		\STATE Variable node $x_c$ computes the message ${f}_{\bar{y}_d \leftarrow x_c}^{\left( t \right) }$ with mean $\mu_{d,c}^{\left( t \right) }$ and variance $\eta_{d,c}^{\left( t \right) }$ by \eqref{15}, then feeds them back to the connected factor nodes;
		\ENDFOR
		\STATE Compute the convergence indicator $\theta^{\left( t \right) }$ in \eqref{16};
		\STATE Update $	\bar{\mathbf{P}} \left( x_c \right)  = {\mathbf{P}}^{\left( t \right) } \left( x_c \right) $, if $\theta^{\left( t \right)} > \theta^{\left( t-1 \right) }$;
		\STATE  $ t = t + 1 $;
	\end{algorithmic}
	\textbf{Until} : $\theta^{\left( t \right) } = 1$ or $t = n_{iter}$; \\
	\textbf{Output}  : The decisions of the transmitted symbols in \eqref{18}.
	\label{alg1}
\end{algorithm} 

\textbf{$\left.  \boldsymbol{1} \right) $ From factor node} $\bar{y}_d$ \textbf{to variable node } $x_c$ : For each factor node, we can represent the signal $\bar{y}_d$ as
\begin{equation}\label{8}
	\bar{y}_d = H_{d,c}x_c +  \underbrace{\sum_{l = 1 }^{c-1} H_{d,l}x_l + \sum_{k=c+1}^{MN} H_{d,k}x_k}_{\text{Interference}} + {\omega}_d .
\end{equation}

The message $f^{\left( t \right) }_{\bar{y}_d \rightarrow x_c} $ from factor node $\bar{y}_d$ to the variable node $x_c$ can be approximated into a Gaussian distribution, i.e.
\begin{subequations}\label{9}
	\begin{align}
		& \log \left(  f^{\left( t \right) }_{\bar{y}_d \rightarrow x_c} \right) \nonumber   \\
		& ~~\propto \text{const} - \frac{\left[ \bar{y}_d - \mathcal{M}_{d,c-1}^{\left( t \right) } - H_{d,c} \left(x_c - {\mu}_{d,c}^{\left( t - 1 \right) } \right) \right]^2 }{ \left( \mathcal{N}_{d,c-1}^{\left( t \right) } - \left|H_{d,c} \right|^2 {\eta}_{d,c}^{\left( t-1 \right) } + \gamma \right)} \\
		& ~~\approx  \text{const}  {- \frac{\left[ \bar{y}_d - \mathcal{M}_{d,c-1}^{\left( t \right) } - H_{d,c} \left(x_c - \hat{\mu}_{c}^{\left( t-1 \right) } \right)   \right]^2 }{ \left( \mathcal{N}_{d,c-1}^{\left( t \right) }  + \gamma \right)  }}, \label{9.3} 
	\end{align}
\end{subequations}
where
\begin{subequations}\label{10}
	\begin{equation}
		\mathcal{M}_{d,c-1}^{\left( t \right) }  = {\sum\nolimits_{l = 1}^{c-1} H_{d,l} \mu_{d,l}^{\left( t \right) }}   +  {\sum\nolimits_{k = c }^{MN} H_{d,k} {\mu}_{d,k}^{\left( t-1 \right) }},
	\end{equation}
	\begin{equation}
		\mathcal{N}_{d,c-1}^{\left( t \right) } = \sum\nolimits_{l=1}^{c-1} \left| H_{d,l} \right|^2  \eta_{d,l}^{\left( t \right) }  +  \sum\nolimits_{k = c }^{MN} \left| H_{d,k} \right|^2  {\eta}_{d,k}^{\left( t-1 \right) },
	\end{equation}
\end{subequations}
$\mu_{d,c}^{\left( t-1 \right) }$ and $\eta_{d,c}^{\left( t-1 \right) }$ are the mean and variance of the message ${f}^{\left( t-1 \right) }_{\bar{y}_d \leftarrow x_c}$ from variable node $x_c$ to factor node $\bar{y}_d$ in $\left( t-1 \right) $-th iteration, $\hat{\mu}_c^{\left( t-1 \right) }$ is the posteriori mean of $c$-th detection symbol in $\left( t-1 \right) $-th iteration. The  basic assumption of AMP framework is applied to \eqref{9.3} with approximation $H_{d,c} \left({{\mu}}_{d,c}^{\left( t \right) } - \hat{\mu}_c^{\left( t \right) } \right) \approx 0 $, $\left|H_{d,c} \right|^2 {\eta}_{d,c}^{\left( t \right) } \approx 0$ for large-scale OTFS systems \cite{c8}. 

However, the connected factor nodes always need to be updated based on the latest feedback extrinsic message ${f}^{\left( t \right) }_{\bar{y}_d \leftarrow x_l}, l = 1, 2, ..., c-1$, leading to a much higher computational load, which motivates employment of the fast recursive scheme to reduce complexity. According to \eqref{10}, we can obtain the following relationships as
\begin{subequations}\label{11.1}
	\begin{equation}
		\mathcal{M}_{d,c}^{\left( t \right) }   =  \mathcal{M}_{d,c-1}^{\left( t \right) } 
		+ H_{d,c}  \left( \mu_{d,c}^{\left( t \right) } - \mu_{d,c}^{\left( t-1 \right) } \right) ,
	\end{equation}
	\begin{equation}
		{\mathcal{N}}_{d,c}^{\left( t \right) }  = \mathcal{N}_{d,c-1}^{\left( t \right) } + \left| H_{d,c} \right| ^2 \left( \eta_{d,c}^{\left( t \right) } - \eta_{d,c}^{\left( t-1 \right) } \right) .
	\end{equation}
\end{subequations}

Therefore, we can apply the simple recursive operations in \eqref{11.1} to substitute the multiple complex multiplications (CMs) in \eqref{10} without any performance-loss. 

Then the message $f^{\left( t \right) }_{\bar{y}_d \rightarrow x_c}$ in \eqref{9.3} can be rearranged as
\begin{align}
		& \log \left(  f^{\left( t \right) }_{\bar{y}_d \rightarrow x_c} \right) \nonumber \\
		& \propto \text{const}  \!-\! \frac{\left|H_{d,c} \right|^2 \nu_{d,c}^s}{2} \left( x_c \!-\! \hat{\mu}_{c}^{\left(t-1\right) } - \frac{H_{d,c} s_{d,c} }{\left|H_{d,c} \right|^2 \nu_{d,c}^s}\right) ^2,
\end{align}
where 
\begin{equation}\label{12}
	\begin{aligned}
		s_{d,c}  =  \frac{\bar{y}_d - \mathcal{M}_{d,c-1}^{\left( t \right) }}{\mathcal{N}_{d,c-1}^{\left( t \right) } + \gamma} ~~,~~ \nu_{d,c}^s  = \frac{1}{\mathcal{N}_{d,c-1}^{\left( t \right) } + \gamma}.
	\end{aligned}
\end{equation}

Therefore, the mean $\chi_{d,c}^{\left( t \right) }$ and variance $ \tau_{d,c}^{\left( t \right) }$ of message $f_{\bar{y}_d \rightarrow x_c}^{\left( t \right) }$ can be given by
\begin{equation}\label{13}
	\chi_{d,c}^{\left( t \right) } = \hat{\mu}_{c}^{\left( t-1 \right) } + \frac{H_{d,c} s_{d,c} }{\left|H_{d,c} \right|^2 \nu_{d,c}^s} ~~, ~~ {\tau}_{d,c}^{\left( t \right) } = \frac{1}{\left|H_{d,c} \right|^2 \nu_{d,c}^s}.
\end{equation}

\textbf{$\left. \boldsymbol{2} \right) $ From variable node} $x_c$ \textbf{to factor node } $\bar{y}_d$ : 
The \emph{a posteriori} probability at variable node $x_c$ can be given by
	\begin{align}
		& {P}^{\left( t \right) }  \left( x_c = \alpha \right) \nonumber \\
		& \propto  {P_{D} \left( x_c = \alpha \right)} \cdot {\prod\limits_{\mathop{e = 1}}^{MN} \exp \left( - \frac{\left| \alpha - \chi_{e,c}^{\left( t \right) } \right|^2 }{\tau_{e,c}^{\left( t \right) }} \right)}, \forall \alpha \in \mathbb{A}, \label{14}
	\end{align}
and we can project it into a Gaussian distribution $\mathcal{CN} \left( x_c; \hat{\mu}_c^{\left( t \right)}, \hat{\eta}_c^{\left( t \right) } \right) $. Therefore, the extrinsic message ${f}_{\bar{y}_d \leftarrow x_c}^{\left( t \right) }$ from variable node $x_c$ to factor node $\bar{y}_d$ can be given by
\begin{subequations}\label{15}
	\begin{equation}
		{\eta}_{d,c}^{\left( t \right) } =   \left[  \left( {\hat{\eta}_c}^{\left( t \right) }\right)^{-1} - \left( {\tau_{d,c}^{\left(t\right) }}\right)^{-1} \right]^{-1} \approx \hat{\eta}_c,
	\end{equation}
	\begin{equation}
		{\mu}_{d,c}^{\left( t \right) } = {\eta}_{d,c}^{\left( t \right) } \left( \frac{\hat{\mu}_c^{\left( t \right) }}{\hat{\eta}_c^{\left( t \right) }} -  \frac{\chi_{d,c}^{\left( t \right) }}{\tau_{d,c}^{\left( t \right) }} \right) = \hat{\mu}_{c}^{\left( t \right) } - \eta_{d,c}^{\left( t \right)}  H_{d,c}s_{d,c}.
	\end{equation}
\end{subequations}

Finally, the variable node $x_c$ feeds the extrinsic message ${f}_{\bar{y}_d \leftarrow x_c}^{\left( t \right) }$ to the connected factor nodes $\bar{y}_d$, $d = 1, 2, ..., MN$ with mean $\mu_{d,c}^{\left( t \right) } $ and variance $\eta_{d,c}^{\left( t \right) }$ for more reliable  interference cancellation of next detection symbol. 

\textbf{$\left. \boldsymbol{3} \right) $ Convergence indicator}: When all variable nodes are updated, we can compute the convergence indicator $\theta^{\left( t \right) } $ by
\begin{equation}\label{16}
	\theta^{\left( t \right) } = \frac{1}{MN} \sum_{c = 1}^{MN} \mathbb{I} \left(  \mathop{\max}\limits_{\alpha \in \mathbb{A}} {{P}^{\left( t \right) } \left( x_c = \alpha \right)} \geq 1 - \varrho \right)
\end{equation}
with a small $\varrho > 0$ and $\mathbb{I} \left( \cdot \right) $ is the indicator function.

\textbf{$\left. \boldsymbol{4} \right) $ Update criterion}: If $\theta^{\left( t \right)} > \theta^{\left( t-1 \right) }$, we can update
\begin{equation}\label{17}
	\bar{\mathbf{P}} \left( x_c \right)  = {\mathbf{P}}^{\left( t \right) } \left( x_c \right), c = 1, 2, ..., MN.
\end{equation}

\textbf{$\left. \boldsymbol{5} \right) $ Stopping criterion:} The UAMP-MFIC iterative detector terminates when either $\theta^{\left( t \right) } = 1$ or the maximum iteration number $n_{iter}$ is reached. Then, we can make the decisions as
\begin{equation}\label{18}
	\hat{x}_c = \mathop{\arg \max}\limits_{\alpha \in \mathbb{A}} \bar{P} \left( x_c = \alpha \right), c = 1, 2, ..., MN.
\end{equation}

From the above algorithm discussion, we can notice that the complexity in the factor nodes is mainly from the fast  recursive scheme in \eqref{11.1}, where the required CMs is given by $\mathcal{O} \left( M^2 N^2 \right) $ for each iteration.  Meanwhile, the complexity in the variable nodes is given by $\mathcal{O} \left(  M N \left| \mathbb{A} \right|  \right) $. Therefore, the overall complexity of proposed UAMP-MFIC detector is given by $\mathcal{O} \left(  M^2 N^2 +   M N \left| \mathbb{A} \right|  \right) $ for each iteration, which is at a similar complexity level with the original UAMP detector.

\section{Proposed Bidirectional Symbol Detectors}

Although the message feedback interference cancellation structure can effectively improve the system performance and achieve fast convergence, it may suffers severe error-propagation from the unreliable estimated symbols and results in performance degradation. The optimal symbol detection order usually requires to apply the enumeration model \cite{c11}, leading to a high computational complexity. In order to obtain a trade-off between complexity and performance, we develop the bidirectional symbol detection structures and respectively consider the forward UAMP-MFIC detection and backward UAMP-MFIC detection without loss of generality, where the detection order is from $x_1$ to $x_{MN}$ and from $x_{MN}$ to $x_{1}$ in forward and backward UAMP-MFIC detectors, respectively. For simplicity, we take the superscript $\left( \cdot \right)^f $ and $\left( \cdot \right)^b $ to represent the corresponding concepts in forward and backward UAMP-MFIC detectors. Then, the T-UAMP-MFIC detector and IW-UAMP-MFIC detector are proposed to efficiently fuse the output information of the forward and backward UAMP-MFIC detectors for alleviating error-propagation and further improving receiver performance. 

\subsection{Turbo UAMP-MFIC Detector}

\begin{figure}[htbp]
	\centering
	\includegraphics[scale=0.40]{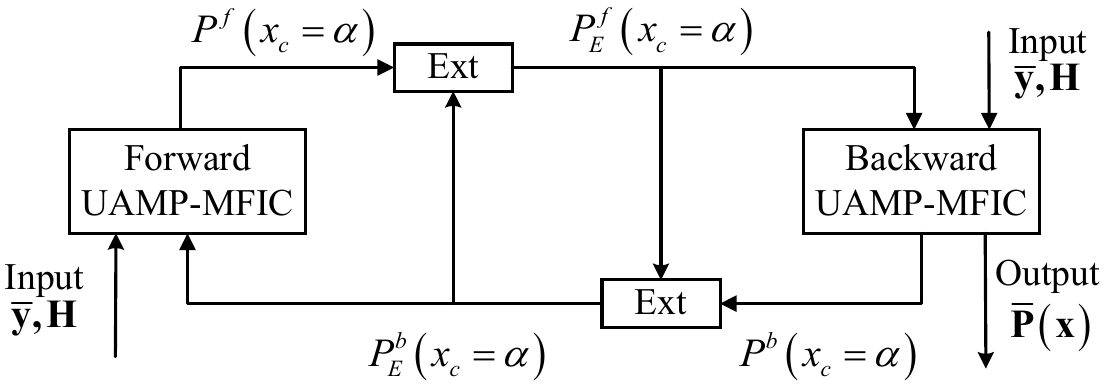}
	\caption{T-UAMP-MFIC detector structure.}
	\label{fig_2}
\end{figure}

The classic Turbo receiver can effectively extract the potential performance gains of equalizers and decoders by exchanging extrinsic information between them to improve detection performance \cite{c12}. Inspired by this, we develop a T-UAMP-MFIC detection structure to enable cooperation of forward and backward UAMP-MFIC detectors, where the details are depicted in Fig. \ref{fig_2}. As iteration continues, the more reliable priori information provided from one can effectively help the other one detector to obtain performance improvement and it will converge after a certain number of iterations. 

Specifically, the forward UAMP-MFIC detector produces {a posteriori} information $ {P}^{f, \left( t \right) }  \left( x_c = \alpha \right) $ on each symbol in $\left( t \right) $-th iteration as
\begin{equation}\label{20}
	\begin{aligned}
		\underbrace{{P}^{f, \left( t \right) }  \left( x_c = \alpha \right)}_{\text{Posteriori}}   \propto   \underbrace{ {P}_{E}^{f, \left( t \right)} \left( x_c = \alpha \right) }_{\text{Extrinsic}} \underbrace{{P}_{D}^{f, \left( t \right)} \left(x_c = \alpha   \right)}_{\text{Priori}}, \forall \alpha \in \mathbb{A},
	\end{aligned}
\end{equation}
where ${P}_{E}^{f, \left( t \right)} \left( x_c = \alpha \right) \propto \prod\limits_{\mathop{e = 1}}^{MN} \exp \left( - \frac{\left| \alpha - \chi_{e,c}^{f, \left( t \right) } \right|^2 }{\tau_{e,c}^{f, \left( t \right) }} \right) $ and
$c = 1, 2, ..., MN$. Then the extrinsic information ${P}_{E}^{f, \left( t \right)} \left( x_c = \alpha \right) $ passes to the backward UAMP-MFIC detector as a priori information, i.e. ${P}_{D}^{b, \left( t \right)} \left( x_c = \alpha\right) =  {P}_{E}^{f, \left( t \right)} \left( x_c = \alpha \right) $. Similarly, the backward UAMP-MFIC detector produces {a posteriori} information ${P}^{b, \left( t \right)} \left( x_c = \alpha \right) $ and passes the extrinsic information ${P}_{E}^{b, \left( t \right)} \left( x_c = \alpha \right) $  back to forward UAMP-MFIC detector as a priori information to form the iterative loop, i.e.  ${P}_{D}^{f,\left( t+1 \right)} \left( x_c = \alpha \right) =  {P}_{E}^{b, \left( t \right)} \left( x_c = \alpha \right) $. Note that, we only pass the extrinsic information during iteration. Otherwise,
the messages will become more and more correlated and the efficiency of iterative detector would be reduced. Since the T-UAMP-MFIC detector involves the iterations between two modules, its detection complexity is given by $\mathcal{O} \left(  2 M^2 N^2 +   2 M N \left| \mathbb{A} \right|  \right) $.  

In order to further alleviate the error-propagation, we also propose a more efficient IW-UAMP-MFIC detector in next sub-section to fully fuse the estimation results of the forward and backward UAMP-MFIC detectors for better performance.

\vspace{-6pt}
\subsection{Iterative Weight UAMP-MFIC Detector}

The posteriori estimation of forward and backward UAMP-MFIC detectors can be respectively expressed as the asymptotic estimation of the original transmitted symbol $x_c$ with an AWGN noise disturbance in $\left( t \right) $-th iteration, i.e.
\vspace{-2pt}
\begin{equation}\label{21}
		\hat{\mu}_c^{f, \left( t \right) }  = x_c + \sqrt{\hat{\eta}_c^{f, \left( t \right) }} z ~,~ \hat{\mu}_c^{b, \left( t \right) }  = x_c + \sqrt{\hat{\eta}_c^{b, \left( t \right) }} z,
\end{equation}
where $z$ is Gaussian variable with distribution $\mathcal{CN} \left( z; 0, 1\right) $. Here, we denote $\varrho^{\left( t \right) } $ as the correlation coefficient of posteriori estimation between the forward and backward detectors, i.e.
\vspace{-2pt}
\begin{equation}\label{23.1}
	\varrho^{\left( t \right) }  = \frac{\mathbb{E} \left[ \hat{\eta}_c^{f, \left( t \right) } \hat{\eta}_c^{b, \left( t \right) } \right] }{ \sqrt{\mathbb{E} \left[ \hat{\eta}_c^{f, \left( t \right) } \right] \mathbb{E} \left[ \hat{\eta}_c^{b, \left( t \right) } \right] } },
\end{equation}
where $\mathbb{E}$ represents the expectation over  $c, c = 1, 2, ..., MN$. We can combine the estimation results of the forward and backward  detectors according to the weighting factor $\lambda_c^{\left( t \right) }$,
\begin{align}
		 \hat{\mu}_c^{w, \left( t \right) }  & = \lambda_c^{\left( t \right) } \hat{\mu}_c^{f, \left( t \right) } + \left( 1 - \lambda_c^{\left( t \right) } \right) \hat{\mu}_c^{b, \left( t \right) }  \nonumber \\
		& = x_c + \lambda_c^{\left( t \right) } \sqrt{\hat{\eta}_c^{f, \left( t \right) }} z + \left( 1 - \lambda_c^{\left( t \right) } \right) \sqrt{\hat{\eta}_c^{b, \left( t \right) }} z.
\end{align}

According to MMSE criteria, the weighting factor $\lambda_c^{\left( t \right) }$ can be optimized as
\vspace{-2pt}
\begin{equation}\label{24}
	\begin{aligned}
		\lambda_c^{\left( t \right) } = \frac{\hat{\eta}_c^{b, \left( t \right)} - \varrho^{\left( t \right)} \sqrt{\hat{\eta}_c^{f, \left( t \right)} \hat{\eta}_c^{b, \left( t \right)}}}{\hat{\eta}_c^{f, \left( t \right)} + \hat{\eta}_c^{b, \left( t \right)} - 2 \varrho^{\left( t \right)} \sqrt{\hat{\eta}_c^{f, \left( t \right)} \hat{\eta}_c^{b, \left( t \right)}}}.
	\end{aligned}
\end{equation}

Finally, since the IW-UAMP-MFIC detector needs to fuse the output information of two modules, its detection complexity is given by $\mathcal{O} \left(  2 M^2 N^2 +   2 M N \left| \mathbb{A} \right|  \right) $ for each iteration, which is similar to that of T-UAMP-MFIC detector.

\vspace{-6pt}
\section{Simulation Results}

In this section, we test the bit error rate (BER) performance of proposed detectors for OTFS systems in high-mobility channels. Without loss of generality, we consider the 16QAM modulated OTFS system with $M = 64$ and $N = 32$. The carrier frequency is 4GHz  with sub-carriers spacing $ \bigtriangleup f =  $ 15kHz. The maximum mobile velocity $ v_{max} = 300\text{km/h} $ and the Doppler shift for each delay is generated by using Jakes’ formulation, i.e. $ v_i = v_{max} \text{cos}(\theta_i) $, where $ \theta_i $ is uniformly distributed over $ \left[ -\pi,\pi \right]  $. The channel coefficient $h_i$ is randomly generated based on a uniform power delay profile. The number of maximum iteration is $n_{{iter}} = 20$. 

\begin{figure*}[t!]
	\captionsetup{font = {scriptsize,scriptsize,scriptsize,small}, format = hang, justification = centering}
	\subfloat[Convergence analysis with different number of iterations.]{\includegraphics[scale=0.39]{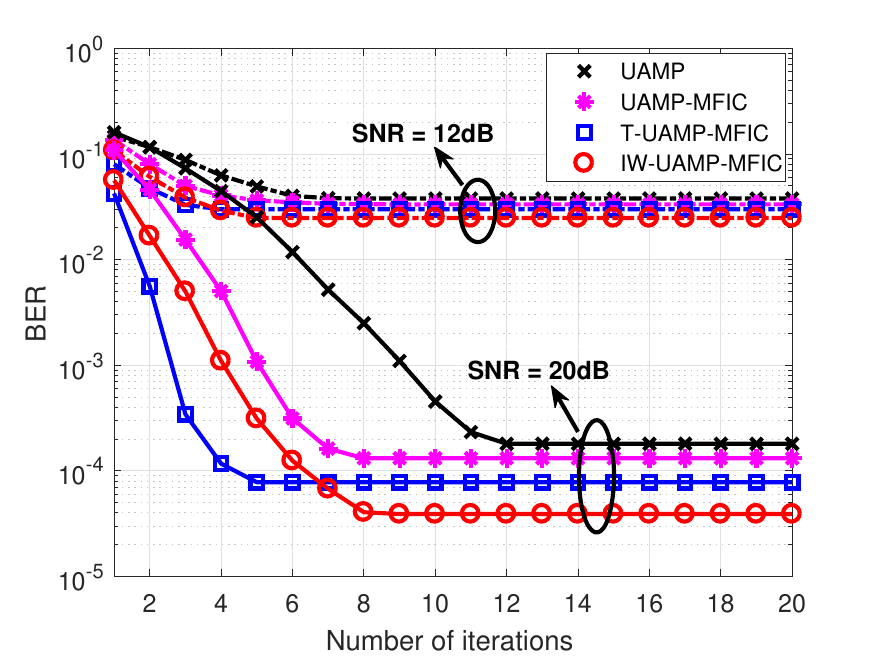}}\label{fig_3.1}
	\hfil
	\subfloat[BER performance with different SNRs.]{\includegraphics[scale=0.39]{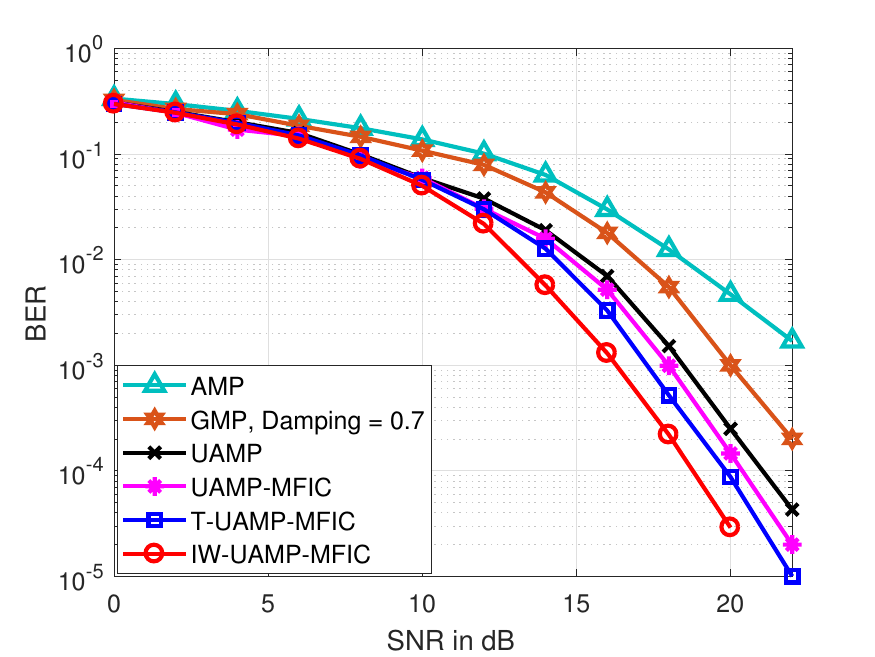}}\label{fig_3.2}
	\hfil
	\subfloat[BER performance with different mobile velocities.]{\includegraphics[scale=0.39]{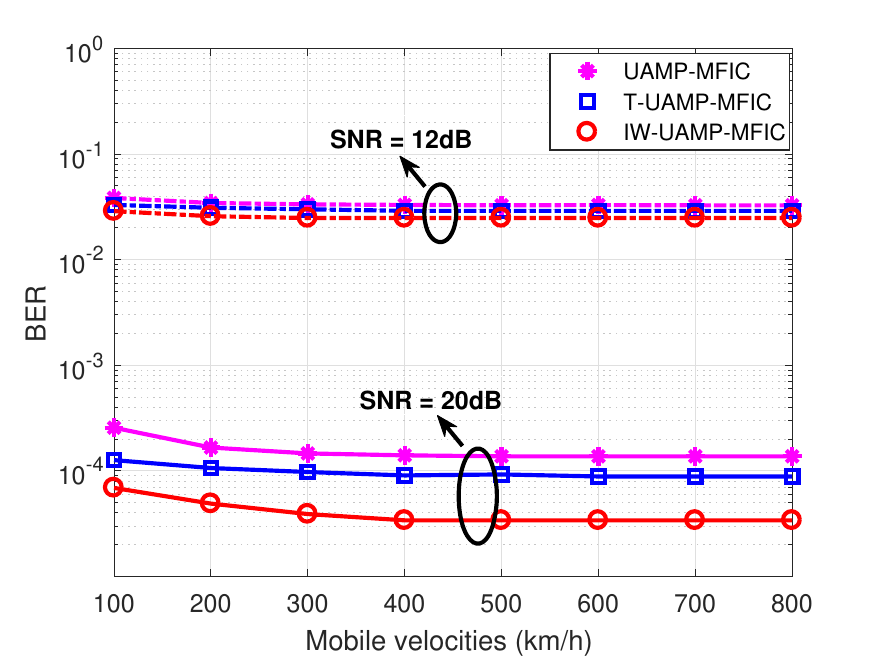}}\label{fig_3.3}
	\hfil
	\vspace{-5pt}
	\caption{Performance analysis of message feedback interference cancellation structure for OTFS systems.}
	\vspace{-15pt}
\end{figure*}

We firstly test the convergence of proposed detectors with different signal-to-noise ratio (SNR) in Fig. 3(a). We can observe that all the detectors converge after a certain number of iteration. The proposed UAMP-MFIC detector shows faster convergence rate and better BER performance than original UAMP detector due to utilize the latest feedback messages for interference cancellation. The proposed T-UAMP-MFIC detector and IW-UAMP-MFIC detector further improve the BER performance of receiver by efficiently fusing the output information of forward and backward UAMP-MFIC detectors to alleviate error-propagation. The IW-UAMP-MFIC detector shows better convergence performance but slower convergence rate than T-UAMP-MFIC detector.

Fig. 3(b) presents the BER performance of proposed detectors with different SNRs. We can notice that all the detectors benefit from higher SNR. The UAMP detector shows better BER performance than GMP \cite{c4}\cite{c5} and AMP \cite{c8} detectors, since the performance of GMP and AMP are limited to specific channel matrix structure. Our proposed UAMP-MFIC detector outperforms the UAMP detector due to more reliable interference cancellation. The T-UAMP-MFIC detector and IW-UAMP-MFIC detector can further improve the BER performance by efficiently fusing the estimation results of forward and backward UAMP-MFIC detectors to alleviate error-propagation. 

Finally, we test the BER performance of proposed detectors at different velocities with SNR = 12dB and 20dB in Fig. 3(c). We can observe that the BER performance improves modestly as velocities growing. The reason is that OTFS modulation can resolve more paths in Doppler dimension with higher velocity and provide higher channel diversity. Consequently, the better performance can be achieved at higher velocities.

\section{Conclusion}

In this letter, we proposed an efficient UAMP-MFIC iterative detector, where the latest feedback extrinsic messages from variable nodes are utilized for more reliable interference cancellation and performance improvement. Then, a fast recursive scheme is leveraged in the proposed UAMP-MFIC detector to prevent complexity increasing. To further alleviate the error-propagation and improve the receiver performance, we also develop the bidirectional symbol detector structures, where T-UAMP-MFIC detector and IW-UAMP-MFIC detector are proposed to efficiently fuse the estimation results of forward and backward UAMP-MFIC detectors. Finally, the simulation results are provided to demonstrate performance improvement of the proposed detectors over existing detectors.

\vfill

\end{document}